\documentstyle[12pt,twoside,fleqn,espcrc1,epsfig]{article}
\newcommand{\pb}{\bar\psi}
\newcommand{\pbp}{\langle\pb\psi\rangle}
\newcommand{\pd}{\psi^\dagger}
\newcommand{\pdp}{\langle\pd\psi\rangle}
\newcommand{\dtp}{\!\frac{d^3p}{(2\pi)^3}}
\def\roughly#1{\mathrel{\raise.3ex\hbox{$#1$\kern-.75em%
\lower1ex\hbox{$\sim$}}}}
\newcommand{\gsim}{\roughly>}

\newcommand{\AmS}{{\protect\the\textfont2
  A\kern-.1667em\lower.5ex\hbox{M}\kern-.125emS}}

\title{Quark Droplets in the NJL Mean Field}

\author{M. Buballa$^{\rm a}$
        and
        M. Oertel\address{Institut f\"ur Kernphysik, 
        TU Darmstadt, \\ 
        Schlo{\ss}gartenstr. 9, D-64289 Darmstadt, Germany}
}

\begin{document}
% typeset front matter
\maketitle

\begin{abstract}
 We study the conditions for the existence of stable quark matter in
 the Nambu--Jona-Lasinio mean field at zero temperature and 
 discuss its interpretation.
\end{abstract}

\section{INTRODUCTION}

Because of its relatively simple structure the Nambu--Jona-Lasinio (NJL)
model \cite{NJL} is one of the most popular models for studying the
spontaneous breaking of chiral symmetry and its restoration at finite
temperatures or densities \cite{Vogl} - \cite{Lutz}. However, most of its 
predictions suffer from the fact that the model is a pure quark model
without confinement, which in general leads to rather unrealistic
scenarios.
In mean field approximation for $T=0$ we can distinguish three different 
cases \cite{MB}:

Case I: 
The phase transition is of second order. Then for all densities there is
a uniform phase of quarks. The quarks are massive at low densities 
and (almost) massless at high densities. The pressure is always
positive and if we do not apply an external force to the system it will
expand to arbitrarily large volumes, i.e. arbitrarily small densities. 

Case II: 
The phase transition is of first order. In this case there will be a regime 
in which two phases coexist. In general, this mixed phase consists of a 
low-density phase ($\rho = \rho_l$) with massive and a high density phase 
($\rho = \rho_h$) with massless quarks. However, for $\rho < \rho_l$ we have 
again a uniform phase of massive quarks. As in case I, low densities are
energetically favored and quark matter is unstable against expansion.

Case III:
The phase transition is of first order, but with $\rho_l=0$. 
Then the low-density component of the mixed phase is the 
vacuum, i.e. we find droplets of massless quarks surrounded by the
non-trivial vacuum.
At least in a schematic way this is very reminiscent of the MIT bag model.
The  droplets are the energetically most favored configuration and 
therefore stable against expansion or collapse. In contrast, any
uniform distribution of quarks at lower densities is unstable against
break-up into the droplet-vacuum phase. 
This scenario was also studied in ref.~\cite{Alford} within
a similar model.

For NJL-like models it depends on the model parameters which of the three
cases is realized. 
However, the non-existence of a uniform dilute gas of quarks in nature
excludes the cases I and II, which both predict such a phase, from being 
realistic. In case III  there is no stable uniform quark gas at low densities
but instead we have the droplet phase. At least if we adopt the interpretation
of quark droplets as schematic bag model baryons this seems to be the 
most realistic scenario of the model. In this article we want to discuss
whether it is also compatible with ``realistic'' model parameters.
To large extent this will be done following ref.~\cite{MB}.

\section{FORMALISM}
 
We consider a Lagrangian for quarks with $n_f=2$ flavors and 
$n_c=3$ colors interacting by NJL-like four-fermion vertices:
\begin{equation}
{\cal L} \;=\; \pb ( i \partial{\hskip-2.0mm}/ - m_0) \psi
            \;+\; G_S [(\pb\psi)^2 + (\pb i\gamma_5{\vec\tau}\psi)^2]
            \;-\; G_V (\pb\gamma^\mu\psi)^2  \ .
\label{L}
\end{equation}
The first two interaction terms (scalar-isoscalar and pseudoscalar-isovector) 
were taken from the original NJL-Lagrangian \cite{NJL}. Since vector
interactions are known to be important at finite densities, e.g. like in the
Walecka model \cite{Walecka}, we have also included a vector-isoscalar term. 
Linearising $\pb\psi$ and $\pd\psi$ about their thermal expectation values
we can calculate the mean field thermodynamic potential (per volume)
at temperature $T$ and chemical potential $\mu$ \cite{AY}. We restrict
ourselves to the Hartree approximation. The result for $T=0$ and $\mu \geq 0$
reads:
\begin{equation}
\omega_{MF}(\mu;\, m, \mu_R) = -12 \int \dtp [E_p
+ (\mu_R-E_p) \, \theta(\mu_R - E_P) ]
\,+\, \frac{(m-m_0)^2}{4G_S} \,-\, \frac{(\mu-\mu_R)^2}{4G_V} \ ,
\label{omf}
\end{equation}
with $E_p = \sqrt{m^2 + {\bf p}^2}$. The integral is strongly divergent
and has to be regularized. For simplicity we use a 3-dimensional sharp
cutoff. The auxiliary variables $m$ (``constituent quark mass'') and $\mu_R$
are defined as $m = m_0 - 2G_S\pbp$ and $\mu_R = \mu - 2G_V\pdp$. 
Since on the other hand the condensates $\pbp$ and $\pdp$ have to be 
calculated from the thermodynamic potential by taking the appropriate 
derivatives we encounter a selfconsistency problem.
It can be shown that it is solved by the 
stationary points of $\omega_{MF}$ with respect to $m$ and $\mu_R$.
This can be used to eliminate $\mu_R$ for given values of $\mu$ and $m$,
$\mu_R = \mu_R(\mu,m)$. Furthermore, we subtract a constant in order to
shift the value of the minimum of the vacuum thermodynamic potential to zero,
i.e. we define:
\begin{equation}
\tilde\omega(\mu,m) \;:=\; \omega_{MF}(\mu;\;m, \mu_R(\mu,m))
\;-\;\omega_{MF}(0;\;m_{vac}, 0) \;.
\label{omb}
\end{equation}
Here $m_{vac}$ is the constituent quark mass which minimizes the 
thermodynamic potential in vacuum. 
For a given value of $\mu$ the extrema of $\tilde\omega$ correspond to the
selfconsistent solutions and the stable solution is given by the absolute
minimum. Other thermodynamic quantities, like the pressure $p$, 
baryon number density $\rho_B = \frac{1}{n_c}\,\pdp$ and energy density 
$\varepsilon$, can be calculated from the thermodynamic potential in the 
standard way.

\section{STABILITY OF QUARK MATTER AT T=0}

With the appropriate choice of parameters chiral symmetry is spontaneously
broken in the NJL-vacuum, i.e. the quarks acquire a constituent mass $m_{vac}$
much larger than the current mass $m_0$.
Unless stated otherwise we will work in the chiral limit, $m_0 = 0$. 
Then the thermodynamic potential $\tilde\omega$ is symmetric about $m=0$.
In vacuum $m=0$ corresponds
to a local maximum and  $\tilde\omega$ is minimal at $m = \pm m_{vac}$.
For large chemical potentials chiral symmetry becomes restored. Thus,
with increasing $\mu$, the maximum at $m=0$ eventually has to turn into an 
absolute minimum. If this happens at a critical chemical potential less 
than the vacuum mass, $\mu_{crit} < m_{vac}$, 
we have a first-order phase transition of the type ``case III''. 

\begin{figure}[htb]
\begin{minipage}[t]{77mm}
{\makebox{\epsfig{file=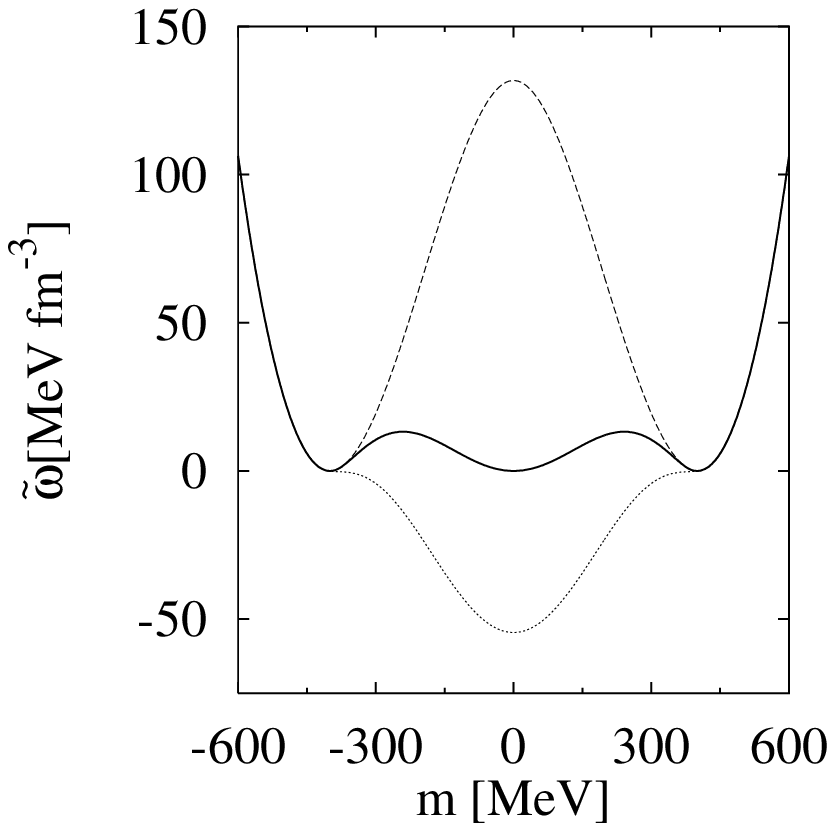,width=68mm}}}
\caption{Thermodynamic potential $\tilde\omega$ as a function of the 
constituent quark mass $m$ for $m_0=0$, $\Lambda$ = 587.9 MeV, 
$G_S\Lambda^2 = 2.48$ and $G_V=0$. The different curves correspond to 
three different chemical potentials:
$\mu=0$ (dashed line), $\mu$ = 376 MeV (solid line) and $\mu$ = 450 MeV
(dotted line).} 
\label{omega}
\end{minipage}
\hspace{\fill}
\begin{minipage}[t]{77mm}
{\makebox{\epsfig{file=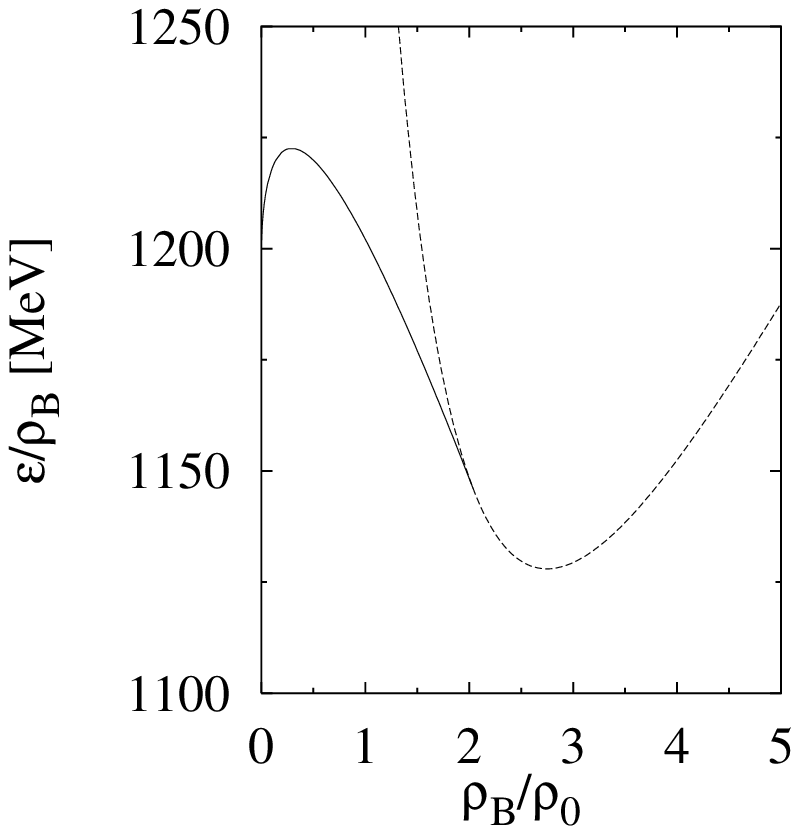,width=68mm}}}
\caption{Energy per baryon number as a function of the
baryon number density $\rho_B$ for the same parameters as in fig.~\ref{omega}.
The solid line corresponds to massive, the dotted line to massless quarks.}
\label{epsilon}
\end{minipage}
\end{figure}

An example for this case is shown in fig.~\ref{omega} where 
we find a critical chemical potential
$\mu_{crit}$ = 376 MeV, well below the vacuum mass $m_{vac}$ = 400 MeV. 
Since for $T=0$ and $G_V=0$ the chemical potential is equal to the Fermi 
energy of the system, for a given mass $m$, a chemical potential less than
$m$ corresponds to zero density and vice versa.  
Hence the degenerate minima of $\tilde\omega$ at $\mu = \mu_{crit}$
correspond to two phases of equal pressure in chemical equilibrium,
one being the nontrivial vacuum ($m=m_{vac}$ and $\rho_B=0$) and the
second consisting of massless quarks at finite density. In our example
we find $\rho_B = 0.47 fm^{-3}= 2.75 \rho_0$ for the dense phase. 
This means, droplets of massless quarks and a baryon number density of 
$2.75 \rho_0$ are stable in vacuum, as far as our mean field 
approach is justified.

Alternatively this can be seen if we calculate the energy per baryon number, 
$\varepsilon/\rho_B$ as a function of density $\rho_B$. The result is
shown in fig.~\ref{epsilon} for the same set of parameters as in
fig.~\ref{omega}. At not too high densities there are solutions with
massive and with massless quarks, and the massive ones are always 
energetically favored. In the limit $\rho_B \rightarrow 0$ the energy per 
baryon number is just 3 times the constituent quark mass. 
However, in agreement with our considerations above we find
the absolute minimum of $\varepsilon/\rho_B$ at $2.75 \rho_0$ which is in  
the regime where only massless solutions exist. At this point the pressure
vanishes and the energy per baryon number is 3 times $\mu_{crit}$. This
shows again that $\mu_{crit}<m_{vac}$ is necessary for this minimum to be 
stable.

The results can become qualitatively different if we change the model
parameters. For instance, with the parameters of fig.~\ref{omega} but with 
a vector coupling $G_V = 0.5 G_S$ we find a first-order phase 
transition at $\mu_{crit}$ = 410.3 MeV $> m_{vac}$. This corresponds to what 
we  called ``case II'' and there is no dense matter solution which can coexist 
with the vacuum. The energy per baryon number as a 
function of density looks similar to fig.~\ref{epsilon}, but with 
the minimum of the massless solutions being only metastable: The corresponding
energy, $3\mu_{crit}$, is now larger than the energy per baryon number for   
$\rho_B \rightarrow 0$, which is $3m_{vac}$. Finally, if we further 
increase the vector coupling, e.g. $G_V = G_S$, we find a second-order 
phase transition and the energy per baryon number is a strictly rising
function of density. 

\begin{figure}[htb]
\begin{minipage}[t]{77mm}
{\makebox{\epsfig{file=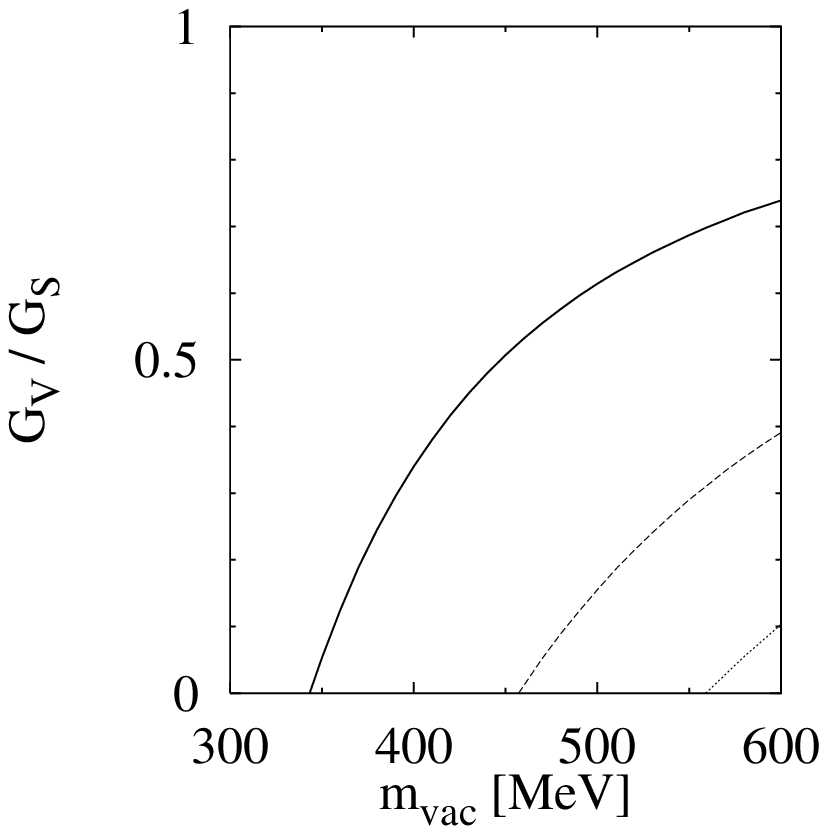,width=68mm}}}
\caption{Lines of constant binding energy per quark for fixed 
 $f_\pi$ = 92.4 MeV and varying constituent quark masses $m_{vac}$ and 
 coupling constants $G_V/G_S$: 0 MeV (solid line), 50 MeV (dashed) 
 and 100 MeV (dotted).} 
 \label{binding}
\end{minipage}
\hspace{\fill}
\begin{minipage}[t]{77mm}
{\makebox{\epsfig{file=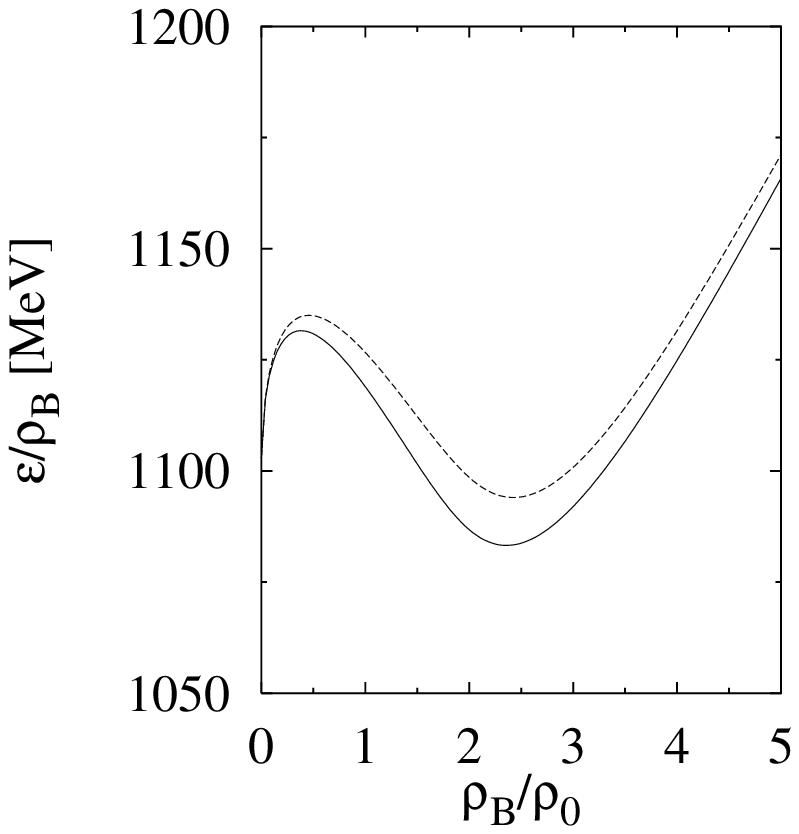,width=68mm}}}
\caption{Energy per baryon number as a function of baryon number density
 for non-strange quark matter in the 3-flavor model. The parameters are:
 $m_0^u=m_0^d$ = 5.5 MeV, $m_0^s$ = 140.7 MeV, $\Lambda$ = 602.3 MeV 
 and $G_S \Lambda^2 = 1.84$, $K\Lambda^5=12.4$ (solid line), 
 $G_S \Lambda^2 = 2.31$, $K=0$ (dashed line).
}
\label{SU3}
\end{minipage}
\end{figure}

A more systematic overview is given in fig.~\ref{binding} where three
lines of constant binding energy per quark, $E_{bind} = m_{vac} -
\mu_{crit}$, are plotted.  For each value of 
$m_{vac}$ the cutoff $\Lambda$ was fixed by fitting the pion decay
constant to its empirical value $f_\pi$ = 92.4 MeV. We are left with
two model parameters, which can be chosen to be $m_{vac}$ and the 
ratio of vector and scalar coupling constant, $G_V/G_S$.   
We see that the vector interaction reduces the binding, which is not
surprising since vector mean fields are known to be repulsive. We
also find that the binding becomes stronger with increasing constituent
quark masses. For $m_{vac}<$ 343 MeV there is no bound quark matter,
even without vector interaction. 
This agrees quite well with an approximate relation derived in ref.~\cite{MB}
which states that stable quark matter is only possible in the NJL-model
if $m_{vac}\gsim 4f_\pi$. 

The region shown in fig.~\ref{binding} corresponds more or less to the
regime of ``realistic parameters''. One could try to further constrain 
$m_{vac}$ by fitting the quark condensate. However, $\pbp$ is not known very 
precisely and its dependence on $m_{vac}$ is only weak, once $f_\pi$ is fixed.
The vector coupling constant could be determined by fitting vector meson
masses. In the literature this usually leads to $G_V/G_S$ of the
order of 0.5 to 1 \cite{Vogl,Lutz} although much higher values are also found
\cite{Ebert}. On the other hand a realistic vector coupling constant
in dense matter could be rather different from the vacuum one.

Thus the scenario of stable quark droplets in vacuum (``case III'') is
possible but not a necessary or very probable consequence of restricting
the model parameters to realistic values. This is in some contradiction
to ref.~\cite{Alford} where stable droplets are found to be most likely.
The model of ref.~\cite{Alford} is almost identical to our model with
$G_V=0$, with the only difference that instead of using a sharp cutoff
the vertices are multiplied with form factors of the form
$(\frac{\Lambda^2}{{\bf p}^2 + \Lambda^2})^\nu$. As an example we looked at
$\nu = 1$ and $\Lambda$ = 800 MeV. It turned out that we are able to produce 
very similar results with our model if we use a sharp cutoff of about
$350$ MeV. However, for this cutoff parameter we find a pion decay constant of 
less than $60$ MeV. This suggests that the parameters chosen in
ref.~\cite{Alford} might also correspond to a too small value of $f_\pi$.

\section{QUARK DROPLETS AS SCHEMATIC MIT-BAGS}

In case III dense matter of massless quarks can coexist with the non-trivial
vacuum. As pointed out in the introduction it is very attractive to 
identify these quark droplets with baryons in a bag-model picture. 
In fact, the energy of a spherical MIT-bag \cite{MIT} with radius $R$,
\begin{equation}
E_{MIT} \;=\; \frac{4\pi}{3} R^3 B \;+\; \frac{3x - z_0}{R}
        \;=\; \frac{B}{\rho_B} \;+\; (3x-z_0)\,
        \left(\frac{4\pi}{3}\right)^{1/3}   \,\rho_B^{1/3}  \ ,
\label{EMIT}
\end{equation}
has a similar structure as the energy per baryon number for massless quarks 
in the NJL mean field
\begin{equation}
\frac{\varepsilon}{\rho_B} \;=\; \frac{B}{\rho_B} \;+\; \frac{3n_c}{4}
        \left(\frac{3\pi^2}{n_f}\right)^{1/3}   \,\rho_B^{1/3}  
         \;+\; G_V \,n_c^2 \,\rho_B   \ ,
\label{ENJL}
\end{equation}
if we switch off the vector interaction. Here the bag constant is given by
$B = \tilde\omega(0,0)$. Of course our thermodynamic approach 
is valid only for a large number of particles in a large volume
whereas eq.~(\ref{EMIT}) was derived for three quarks in a volume
which can be small. This leads to different coefficients in front of
$\rho_B^{1/3}$. However, taking $x=2.04$ and $z_0=1.84$ \cite{MIT},
the deviation turns out to be surprisingly small ($\sim 20\%$).
One should also keep in mind that eq.~(\ref{ENJL}) is only valid for the 
massless quark solutions. As we have seen, if the density is low enough the 
quarks can lower their energy by acquiring a finite constituent mass (cf. 
fig.~\ref{epsilon}). In particular, only a finite amount of energy is
needed to lower the density to zero, corresponding to an infinite bag
radius: the NJL mean field does not confine. As an important consequence
the quark droplets become immediately unstable against evaporation 
of massive quarks at any finite temperature and one has to be very careful 
with the interpretation of the model for $T>0$.

\section{THREE-FLAVOR MODEL}

In this section we extend the model to three quark flavors
\cite{Klimt}:
\begin{eqnarray}
{\cal L} \;=\; \pb ( i \partial{\hskip-2.0mm}/ - {\hat m_0}) \psi
            \;&+&\; G_S \sum_{k=0}^8 [\,(\pb\lambda_k\psi)^2 + 
           (\pb i\gamma_5\lambda_k\psi)^2\,] \nonumber \\
            \;&-&\; K \,[ \,det_f (\pb(1+\gamma_5)\psi) 
                         + det_f (\pb(1-\gamma_5)\psi) \,]   \ .
\label{L3}
\end{eqnarray}
Here $\psi = (u,d,s)^T$ is a 3-dimensional vector and 
${\hat m_0} = diag(m_0^u,m_0^d,m_0^s)$ a $3 \times 3$ matrix in flavor space. 
In addition to the 4-point interaction 
the Lagrangian contains a t'Hooft-type 6-point interaction 
which is a determinant in flavor space and breaks the $U_A(1)$
symmetry. For simplicity we neglect vector interactions in this section.
Since each flavor is separately conserved we have in principle three
different chemical potentials and the thermodynamic potential contains
the term 
$\mu N \equiv \mu_u u^\dagger u + \mu_d d^\dagger d + \mu_s s^\dagger s$.
\footnote{Here we use the notation of ref.~\cite{Lutz}. Notice that 
$\mu_s$ is different from the strangeness chemical potential
$\mu_{strange}$ which is often introduced as the deviation from a common baryon
chemical potential $\mu_B$:  
$\mu N = \mu_B (u^\dagger u + d^\dagger d + s^\dagger s) +
\mu_{strange} s^\dagger s$.
In particular in our notation $\mu_s=0$ corresponds to
$\langle s^\dagger s \rangle = 0$.} 

In the following we consider non-strange isospin-symmetric matter,
$m_0^u=m_0^d$, $\mu_u=\mu_d$ and $\mu_s=0$.      
We adopt the model parameters of ref.~\cite{Rehberg} which have been fitted
to the pseudoscalar mass spectrum and $f_\pi$.
For these parameters we find stable quark matter with
$\rho = 2.36\rho_0$ and a binding energy per quark of $6.5$ MeV. 
The energy per baryon number is plotted in fig.~\ref{SU3} (solid line). 
Because of the determinant interaction this result is influenced by the 
strange quark
even though the net strangeness $\langle s^\dagger s\rangle$ is zero.
To show this we switch off the determinant interaction 
and increase $G_S$ such that the vacuum properties of the
non-strange sector, $\langle \bar u u\rangle$, $f_\pi$ and $m_\pi$, 
remain unchanged. The result is also shown 
in fig.~\ref{SU3} (dashed line). Here the binding energy is $2.9$ MeV at 
$\rho = 2.43\rho_0$. Thus the flavor mixing gives some extra binding but the 
effect is not very large. A more systematic discussion of the $SU(3)$ model
will be published elsewhere.

\end{document}